\documentstyle [12pt] {article}

\catcode`@=12
\newcommand{\be}{\begin{equation}}
\newcommand{\ee}{\end{equation}}
\newcommand{\ber}{\begin{eqnarray}}
\newcommand{\eer}{\end{eqnarray}}
\newcommand{\bers}{\begin{eqnarray*}}
\newcommand{\eers}{\end{eqnarray*}}
\begin{document}
\vspace{0.5in}
\oddsidemargin -.375in  
\newcount\sectionnumber 
\sectionnumber=0 
\vspace {.5in} 
\thispagestyle{empty}
\begin{flushright} UH-511-864-97 \\
ISU-HET-97-07\\ July 1997\\
\end{flushright}
\vspace {.5in}
\begin{center}
{ \bf{Quasi-Inclusive and Exclusive decays of  $B $ to $\eta'$ \\}
\vspace{.5in}
{\rm {A.
Datta{\footnote{email:
datta@iastate.edu, datta@medb.physics.utoronto.ca}
}$^a$,
X.-G. He{\footnote{
hexg@dirac.ph.unimelb.edu.au}
}and S. Pakvasa{\footnote
{pakvasa@uhheph.phys.hawaii.edu}
}} \\}

\vskip .3in
{\it $^1 $Department of Physics and Astronomy \\}
{\it  Iowa State University, \\}
{\it Ames, Iowa 50011\\}

{\it $^{a} $Department of Physics \\}
{\it University of Toronto, \\}
{\it Toronto, Ontario, Canada M5S 1A7\\}

{\it $^2$ School of Physics\\}
{\it  University of Melbourne, \\}
{\it Parkville, Victoria, 3052 \\}
{\it Australia\\}

{\it And\\}
{\it $^3$Department of Physics and Astronomy \\}
{\it University of Hawaii, \\}
{\it Honolulu, Hawaii 96822\\}

}
\vskip .5in
\end{center}  
\begin{abstract}
We consider the effective Hamiltonian of four quark operators in the
Standard Model
in the exclusive and  quasi-inclusive   
decays of the type $B\rightarrow K^{(*)} \eta^{\prime}$, $B\rightarrow \eta' X_s$,
 where $X_s$  contains a single Kaon. Working in the factorization
assumption we find that the four quark operators can account for the 
recently measured 
exclusive decays $B\to \eta^{\prime}(\eta) K$ and $B\to K \pi$ for appropriate choice of form factors
but cannot explain the large quasi-inclusive rate.
\end {abstract}
\baselineskip 24pt

%

%
%

\section{Introduction}
Recently CLEO has reported
a branching ratio for the process
$B^- \to \eta' K^- $  at 
$(7.1 ^{+2.5}_{-2.1} \pm 0.9)\times 10^{-5}$ \cite{CLEO1}. Upper limits
on related exclusive 
decay modes with $\eta$ or $\eta^{'}$ with a $K^*$ mesons in the final
states
 were also obtained. 
   The quasi-inclusive process 
$B \to \eta' X_s$ for high momentum $\eta'$ was also measured
with a high rate
\cite{CLEO2}
\ber
B \to \eta' X_s & =&(6.2\pm 1.6\pm 1.3)\times 10^{-4}
(p_{\eta'}>2.0GeV)\
\eer
where $X_s$ stands for one Kaon and up to 4 pions with at most one
$\pi^0$.

Several explanations for the large quasi-inclusive decay rate have been
proposed both within the Standard Model and beyond \cite{AS,ZHIT,HT,KUAN}. 
However, some of these
analyses have underestimated the effect of the effective four quark
operators to this process. For example,
it has been assumed that this contribution is
small based on the smallness of $V_{ub}$\cite{ZHIT}. 
However, the penguin contribution to
this process cannot be neglected. A simple estimate based on the
magnitudes of the CKM elements associated with the tree and the penguin
diagrams clearly suggests that this process is dominated by penguin
contributions if one considers only the four quark effective
Hamiltonian. 
In this paper we calculate the contribution of the
effective Hamiltonian of the four quark operators to the exclusive
decays $ B \to K^{(*)} \eta^{\prime}(\eta)$ and $ B \to K \pi $
along with the
 quasi-inclusive decay of the $B \to \eta' X_s$.

In the sections which follow, we describe the effective Hamiltonian of
four quark operators, our calculation of the the exclusive and
quasi-inclusive rates.

\section{ Effective Hamiltonian} 
In the Standard Model (SM) 
the amplitudes for hadronic $B$ decays of the type $b\to q \bar{f} f$ 
are generated by the following effective 
Hamiltonian \cite{Reina}:
\begin{eqnarray}
H_{eff}^q &=& {G_F \over \protect \sqrt{2}} 
   [V_{fb}V^*_{fq}(c_1O_{1f}^q + c_2 O_{2f}^q) -
     \sum_{i=3}^{10}(V_{ub}V^*_{uq} c_i^u
+V_{cb}V^*_{cq} c_i^c +V_{tb}V^*_{tq} c_i^t) O_i^q] +H.C.\;
\end{eqnarray}
where the
superscript $u,\;c,\;t$ indicates the internal quark, $f$ can be $u$ or 
$c$ quark and $q$ can be either a $d$ or a $s$ quark depending on 
whether the decay is a $\Delta S = 0$
or $\Delta S = -1$ process.
The operators $O_i^q$ are defined as
\begin{eqnarray}
O_{f1}^q &=& \bar q_\alpha \gamma_\mu Lf_\beta\bar
f_\beta\gamma^\mu Lb_\alpha\;\;\;\;\;\;\;O_{2f}^q =\bar q
\gamma_\mu L f\bar
f\gamma^\mu L b\;\nonumber\\
O_{3,5}^q &=&\bar q \gamma_\mu L b
\bar q' \gamma_\mu L(R) q'\;\;\;\;\;\;\;\;O_{4,6}^q = \bar q_\alpha
\gamma_\mu Lb_\beta
\bar q'_\beta \gamma_\mu L(R) q'_\alpha\;\\
O_{7,9}^q &=& {3\over 2}\bar q \gamma_\mu L b  e_{q'}\bar q'
\gamma^\mu R(L)q'\;\;O_{8,10}^q = {3\over 2}\bar q_\alpha
\gamma_\mu L b_\beta
e_{q'}\bar q'_\beta \gamma_\mu R(L) q'_\alpha\;\nonumber
\end{eqnarray}
where $R(L) = 1 \pm \gamma_5$, 
and $q'$ is summed over u, d, and s.  $O_1$ are the tree
level and QCD corrected operators. $O_{3-6}$ are the strong gluon induced
penguin operators, and operators 
$O_{7-10}$ are due to $\gamma$ and Z exchange (electroweak penguins),
and ``box'' diagrams at loop level. The Wilson coefficients
 $c_i^f$ are defined at the scale $\mu \approx m_b$ 
and have been evaluated to next-to-leading order in QCD.
The $c^t_i$ are the regularization scheme 
independent values obtained in Ref. \cite{FSHe}.
We give the non-zero  $c_i^f$ 
below for $m_t = 176$ GeV, $\alpha_s(m_Z) = 0.117$,
and $\mu = m_b = 5$ GeV,
\begin{eqnarray}
c_1 &=& -0.307\;,\;\; c_2 = 1.147\;,\;\;
c^t_3 =0.017\;,\;\; c^t_4 =-0.037\;,\;\;
c^t_5 =0.010\;,
 c^t_6 =-0.045\;,\nonumber\\
c^t_7 &=&-1.24\times 10^{-5}\;,\;\; c_8^t = 3.77\times 10^{-4}\;,\;\;
c_9^t =-0.010\;,\;\; c_{10}^t =2.06\times 10^{-3}\;, \nonumber\\
c_{3,5}^{u,c} &=& -c_{4,6}^{u,c}/N_c = P^{u,c}_s/N_c\;,\;\;
c_{7,9}^{u,c} = P^{u,c}_e\;,\;\; c_{8,10}^{u,c} = 0
\end{eqnarray}
where $N_c$ is the number of color. 
The leading contributions to $P^i_{s,e}$ are given by:
 $P^i_s = ({\frac{\alpha_s}{8\pi}}) c_2 ({\frac{10}{9}} +G(m_i,\mu,q^2))$ and
$P^i_e = ({\frac{\alpha_{em}}{9\pi}})
(N_c c_1+ c_2) ({\frac{10}{9}} + G(m_i,\mu,q^2))$.  
The function
$G(m,\mu,q^2)$ is given by
\begin{eqnarray}
G(m,\mu,q^2) = 4\int^1_0 x(1-x)  \mbox{ln}{m^2-x(1-x)q^2\over
\mu^2} ~\mbox{d}x \;
\end{eqnarray}
All the above coefficients are obtained up to one loop order in electroweak 
interactions. The momentum $q$ is the momentum carried by the virtual gluon in
the penguin diagram.
When $q^2 > 4m^2$, $G(m,\mu,q^2)$ develops an imaginary part. 
In our calculation, we 
use $m_u = 5$ MeV, $m_d = 7$ MeV, $m_s = 200$ MeV, $m_c = 1.35$ GeV
\cite{lg,PRD} and use $q^2=m_b^2/2$.

\section{Matrix Elements for $ B  \rightarrow {K^{(*)}}  \eta^{\prime}(\eta) $,
 $B \to K \pi $
and $B  \rightarrow X_s  \eta^{\prime} $}

 The effective Hamiltonian described in the previous section
 consists of operators with a current $\times$ current structure. 
Pairs of such operators can be expressed in terms of
color singlet and color octet structures which lead to color singlet and
color octet matrix elements. We use the factorization approximation, where
one separates out the currents in the operators by inserting the vacuum
state and neglecting any QCD interactions between the two currents. The
basis for this approximation is that, if the quark pair created by one of
the currents carries large energy then it will not have significant 
QCD interactions. Factorization appears to describe nonleptonic B decays
rather well\cite{SN} . To accommodate some 
deviation from this approximation one can treat
$N_c$, the number of colors that enter in the calculation of the matrix
elements, as a free parameter though the  value of $N_c\sim 2$ is
suggested by experimental data on 
low multiplicity hadronic $B$ decays.
In this section we describe the calculation of matrix elements for the
exclusive decays $B\rightarrow \eta^{\prime}(\eta) K^{(*)}$, $ B \to K \pi $ and 
the quasi-inclusive decay
$B\rightarrow \eta^\prime X_s$.

The $\eta$ and $\eta^\prime$ mesons are mixtures of singlet and octet states
$\eta_1$ and $\eta_8$ of $SU(3)$.
\ber
	\left[ \begin{array}{c} \eta \\ \eta' \end{array} \right]
	&=& \left[
		\begin{array}{cc}
		\cos \theta & - \sin \theta \\
		\sin \theta & \cos \theta
		\end{array}
	  \right]
	  \left[
		\begin{array}{c}
		\eta_8 \\ \eta_1
		\end{array}
	  \right] \\
	\eta_8 &=& \frac{1}{\sqrt{6}} \left[ u \bar{u} +
		d \bar{d} -2 s \bar{s} \right] \\
	\eta_1 &=& \frac{1}{\sqrt{3}} \left[ u \bar{u} +
		d \bar{d} + s \bar{s} \right]
\eer
where the mixing angle $\theta$ lies between $-10^0$ and $-20^0$\cite{PRD}.
We can express the decay constants $f_{\eta^\prime}^u,
f_{\eta^\prime}^d, f_{\eta^\prime}^s$ in terms of octet and singlet
decay constants $f_1$,$f_8$
\ber
	f_{\eta^\prime}^u = f_{\eta^\prime}^d
		&=& \sqrt {\frac{1}{3}} f_\pi
		\left[ \cos\theta \frac{f_1}{f_\pi}
			+ \frac {\sin\theta}{\sqrt{2}} \frac{f_8}{f_\pi}
		\right] \\
	f_{\eta^\prime}^s &=& \sqrt {\frac{1}{3}} f_\pi
		\left[ \cos\theta \frac{f_1}{f_\pi}
			- \sqrt{2} \sin\theta \frac{f_8}{f_\pi}
		\right]\; \
\eer
and
similarly one has
\ber
	f_\eta^u =  f_\eta^d &=& \sqrt {\frac{1}{3}} f_\pi
		\left[
		\frac{1}{\sqrt{2}} \cos\theta \frac{f_8}{f_\pi}
		- \sin\theta \frac{f_1}{f_\pi}
		\right] \\
	f_\eta^s &=& - \sqrt {\frac{1}{3}} f_\pi
		\left[
		\sqrt{2} \cos\theta \frac{f_8}{f_\pi}
		+ \sin\theta \frac{f_1}{f_\pi}
		\right] \
\eer
In the $SU(3)$ and Nonet symmetry limit, the relations
$f_8 = f_\pi =130$ MeV and $f_1 = f_\pi$
hold. However these symmetries are not exact and we will consider values
for $f_1$ and$f_8$ away from this symmetry limit. For example the
value of $f_K\sim 1.28f_{\pi}$ indicates about a 30\% SU(3) breaking effect.
\subsection{Exclusive Decay}
In this section we calculate the rates for the two body noleptonic decay
for $B\to \eta^{\prime}(\eta) K$ and $B\to \pi K$. 

To evaluate the rates for the exclusive decays we have to calculate matrix
element of the type $<h K| H_{eff} | B>$ where $h$ is a $\pi$ or
$\eta^{\prime}(\eta)$ and 
 $H_{eff}$, as
already described in the previous section, has the form.
\be
	H_{eff} = V_u (c_1 O_1 + c_2 O_2)
		- \sum_{i=3}^{10}
		\left[ V_u c_i^u + V_c c_i^c + V_t c_i^t \right] O_i
\ee
with $V_u = V_{us}^{*} V_{ub}$, 
	$V_c = V_{cs}^{*} V_{cb}$ and
	$V_t = V_{ts}^{*} V_{tb}$.

As an input to our calculation we need the form factors defined through
\ber
	<K(p_K)|\bar{s} \gamma_\mu (1-\gamma_5) b | B(p_B) >
	&=& \left[ (p_B+ p_K)_\mu - \frac{m_B^2-m_K^2}{q^2} q_\mu \right]
	  F_1^K(q^2)\nonumber\\
	 & + & \frac{m_B^2-m_K^2}{q^2} q_\mu F_0^K(q^2) \
\eer
where $q = p_B -p_K $.

The matrix element
$ < h | \bar{u} \gamma_\mu (1-\gamma_5) b | B > $
is similarly described in terms of form factors
$F_1^{h}(q^2)$ and  $F_0^{h}(q^2)$.

For the $K^{*}$ modes we define
\be
	<K^{*\lambda} | \bar{s} \gamma_\mu (1-\gamma_5) u | 0 >
	= m_{K^{*}} g_{K^{*}} \varepsilon_\mu^{\lambda *} \
\ee
The form factors in this case are defined as
\ber
<K^{*}(p')|J^{\mu}|B(p)>
& = & \frac{2V(k^2)}{m_B + m_{K^*}} \varepsilon^{\mu \nu \alpha
\beta}\varepsilon_{\nu}^{*}p_{\alpha}
p'_{\beta}
+ i(m_B + m_{K^*}) A_1(k^2) \varepsilon _{}^{*\mu}  
-i \varepsilon^{*} \cdot k\frac{A_2(k^2)}{m_B +m_{K^*}} (p+p')^{\mu}\nonumber\\
&+&i \varepsilon^{*} \cdot k \frac{2m_{K^*} (A_0(k^2) -
A_3(k^2))}{k^2}k^{\mu}\
\eer
with
\bers
A_3(k^2) &=& \frac{(m_B + m_{K^*})A_1 - (m_B - m_{K^*})A_2}
{2m_{K^*}}\
\eers
where $k=p_B-p_{K^*}$.
\subsubsection {\bf $B^{\pm}\to K^{\pm} \eta^{\prime}(\eta)$}

The amplitude for $ B^- \to K^- \eta'$ can be written as
\ber
        M & = & \frac{G_F}{\sqrt 2}
        \left[V_u \left(a_1 r_1 Q_K + a_2 Q_{\eta'} \right)
        -\sum_{i=u,c,t}V_i \left\{(T_1^i r_1 +T_2^i r_2)Q_K +T_3^i
Q_{\eta'}\right\}\right]\
\eer
where
\bers
         T_1^i &= & 2a_3^i- 2a_5^i- \frac{1}{2}a_7^i + \frac{1}{2}a_9^i\\
         T_2^i & = & a_3^i +a_4^i- a_5^i
                   +(2a_6^i -a_8^i)
\frac{m_{\eta^\prime}^2}{2m_s(m_b-m_s)} + \frac{1}{2}a_7^i 
                    - \frac{1}{2}a_9^i - \frac{1}{2}a_{10}^i\\
         T_3^i & = & a_4^i + 2(a_6^i +a_8^i)
\frac{m_K^2}{m_u+m_s} \frac{1}{m_b-m_u} 
 +a_{10}^i \
\eers
with 
$a_1  =  c_1 + \frac{c_2}{N_c} $,
$        a_2 = c_2 + \frac{c_1}{N_c} $,
        $a^i_j  =  c^i_j + \frac{c^i_{j+1}}{N_c} $, 
        $a^i_{j+1} =  c^i_{j+1} + \frac{c^i_j}{N_c}$, 
	$r_1 = \frac{f_{\eta^\prime}^u}{f_\pi}$
	$r_2 = \frac{f_{\eta^\prime}^s}{f_\pi} $,
	$Q_K = i F_0^K (m_{\eta^\prime}^2) (m_B^2 - m_K^2) f_\pi$,
	$Q_{\eta^\prime} = i F_0^{\eta^\prime} (m_K^2)
		(m_B^2 - m_{\eta^\prime}^2) f_K $,	
$V_i = V_u, V_c, V_t$ and $N_c$ is effective number of colors.

In the above equations we have used the quark equations of motion to 
simplify certain matrix
elements. The expression for the amplitude can also be used for
$B\rightarrow \eta K$ by making the necessary changes. It is also
straight forward to write down the amplitudes for
$B\rightarrow K^{*} \eta^\prime$ and $B\rightarrow K^{*} \eta$ decays.

\ber
        M & = & \frac{G_F}{\sqrt 2}
        \left[V_u \left(a_1 f_{\eta'}^u A + a_2 m_{K^*}g_{K^*} B \right)
        -\sum_{i=u,c,t}V_i \left\{(T_1^i f_{\eta'}^u +
T_2^i f_{\eta'}^s )A +T_3^i
m_{K^*}g_{K^*} B\right\}\right]\
\eer
where
\bers
         T_1^i &= & 2a_3^i- 2a_5^i- \frac{1}{2}a_7^i + \frac{1}{2}a_9^i\\
         T_2^i & = & a_3^i +a_4^i- a_5^i
                   -(2a_6^i -a_8^i)
\frac{m_{\eta^\prime}^2}
		{2 m_s (m_b + m_s)} + \frac{1}{2}a_7^i 
                    - \frac{1}{2}a_9^i - \frac{1}{2}a_{10}^i\\
         T_3^i & = & a_4^i  +a_{10}^i \
\eers

with
	$A = 2m_{K^*}A_0 \varepsilon^*\cdot p_B $,
	$B = 2 \varepsilon^{ *} \cdot p_B
		F_1^{\eta^\prime} (m_{K^*}^2)$ 
and we will use $g_{K^{*}} =221 $ MeV.
A similar expression for $B\rightarrow \eta K^{*}$ can also be obtained.

\subsubsection {\bf $B\to K  \pi$}
We give below expressions for the various $B\to K  \pi$ decays

\ber
M(B^-\to K  \pi^-) & = &-\frac{G_F}{\sqrt{2}}
V_i\left[ a_4^i -\frac{1}{2}a_{10}^i +\frac{2m_K^2}{(m_s
+m_d)(m_b-m_d)}(a_6^i-\frac{1}{2}a_8^i)\right]Q_{\pi}\nonumber\\
Q_{\pi} & = & if_K(m_B^2-m_{\pi}^2)F_0^{B \to \pi}(m_K^2)\nonumber\\
M(B^-\to K^*  \pi^-) & = &-\frac{G_F}{\sqrt{2}}
V_i\left[ a_4^i -\frac{1}{2}a_{10}^i \right]
2m_{K^*}f_{K^*}\varepsilon^*.p_BF_1^{B \to \pi}(m_{K^*}^2)\
\eer
\ber
M({\bar B}^0\to K^-  \pi^+) & = &\frac{G_F}{\sqrt{2}}
\left [V_ua_2 -V_i\left\{ a_4^i +a_{10}^i +\frac{2m_K^2}{(m_s
+m_u)(m_b-m_u)}(a_6^i+a_8^i)\right\}\right]Q_{\pi}\nonumber\\
Q_{\pi} & = & if_K(m_B^2-m_{\pi}^2)F_0^{B \to \pi}(m_K^2)\nonumber\\
M({\bar B}^0\to K^{-*}  \pi^+) & = &\frac{G_F}{\sqrt{2}}
\left [V_ua_2 -V_i\left\{ a_4^i +a_{10}^i \right\}\right]
 2m_{K^*}f_{K^*}\varepsilon^*.p_BF_1^{B \to \pi}(m_{K^*}^2)\
\eer
\ber
M(B^-\to K^-  \pi^0) & = &\frac{G_F}{\sqrt{2}}
\left[V_u\left\{ a_1Q_K +a_2Q_{\pi}\right\} -V_i
\left \{ T_1^iQ_K +T_2^iQ_{\pi}\right\}\right]\nonumber\\
M(B^-\to K^{*-}  \pi^0) & = &\frac{G_F}{\sqrt{2}}
\left[V_u\left\{ a_1Q_{K'} +a_2Q_{{\pi}'}\right\} -V_i
\left \{ T_1^{i'}Q_{K'} +T_2^{i'}Q_{{\pi}'}\right\}\right]
\eer
where
$T_1^i = -\frac{3}{2}a_9^i +\frac{3}{2}a_9^i =T_1^{i'}$,
$T_2^i = a_4^i +a_{10}^i +\frac{2m_K^2}{(m_s
+m_u)(m_b-m_u)}(a_6^i+a_8^i)$, $T_2^{i'} = a_4^i +a_{10}^i $,
$Q_{\pi} =  if_K(m_B^2-m_{\pi}^2)F_0^{B \to \pi^0}(m_K^2)$ and
$Q_{K}  =  i\frac{f_{\pi}}{\sqrt2}(m_B^2-m_{K}^2)F_0^{B \to K}(m_{\pi}^2)$
$Q_{{\pi}'}  =  2m_{K^*}f_{K^*}\varepsilon^*.p_BF_1^{B \to \pi^0} $, and
$Q_{K'}  =  2m_{K^*}\varepsilon^*.p_BA_0^{B\to K^*}\frac{f_{\pi}}{\sqrt
2}$
\subsection{Quasi-inclusive Decay}
The technique to handle quasi-inclusive decays have been described
in detail in Ref.\cite{BDXP}. We represent the total amplitude
as the sum of a
 ``two body" and a ``three body" piece. The ``two body" amplitude
is given by
\ber
	M_2 &=& i \frac {G_F}{\sqrt{2}} f_{\eta^\prime}^u p_\mu^{\eta^\prime}
		<X| \bar{s} \gamma^\mu (1-\gamma_5) b | B > FL_u \nonumber\\
	& &	- i \frac {G_F}{\sqrt{2}}
		<X|\bar{s} \gamma^\mu (1-\gamma_5) b | B >
		\frac {m_{\eta^\prime}^2} {2 m_s} r
		f_{\eta^\prime}^u FR_u
\eer
where
\bers
	FL_u &=& V_u \left( c_1 + \frac {c_2}{N_c} \right)
		+ A_3 \left\{ 2 + \left(  1+ \frac{1}{N_c} \right)
		r \right\} + A_4 
		\left\{ \frac{2}{N_c} + \left( \frac{1}{N_c} + 1 \right)
		r \right\} \\
	& &	+ (-2-r)\left( A_5+ \frac {A_6}{N_c} \right)
		- \frac{1}{2} \left( 1-r \right)
		\left( A_7+\frac{A_8}{N_c} \right) \\
	& &	+ \frac{1}{2} \left\{ 1-r
		\left( \frac{1}{N_c} + 1\right) \right\} A_9
		+ \frac{1}{2} \left\{ \frac{1}{N_c} - \left(
		\frac{1}{N_c}+1\right) r \right\}
		A_{10} \nonumber\\
	FR_u &=& -2 \left( \frac{A_5}{N_c}+A_6\right)
		+ \left( \frac{A_7}{N_c}+A_8\right)
\eers
with $A_i  = -(c_i^u V_u + c_i^c V_c + c_i^t V_t) $ and
	$r  = \frac {f_{\eta^\prime}^s} {f_{\eta^\prime}^u}$

The ``three body" piece has the form
\ber
	M_3 &=& <\eta | \bar{u} \gamma^\mu (1-\gamma_5) b | B >
		<X| \bar{s} \gamma_\mu (1-\gamma_5) u | 0 > L_u \nonumber\\
	& &	+ <\eta | \bar{u} (1-\gamma_5) b | B >
		<X|\bar{s}(1+\gamma_5) u | 0 > E_u \
\eer
with
\bers
	L_u &=& \left( \frac{c_1}{N_c} + c_2 \right) V_u
		+ \left( \frac{A_3}{N_c} + A_4 \right)
		+ \left( \frac{A_9}{N_c} + A_{10} \right) \\
	E_u &=& -2 \left( \frac{A_5}{N_c} + A_6 \right)
		-2 \left( \frac{A_7}{N_c} + A_8 \right)
\eers
Details of the calculations of the branching fraction and decay distributions
are given in Ref \cite{BDXP}.

\section{Results and Discussion }
For the exclusive decays to the $\eta'$ the inputs to the calculation 
are the form
factors, the values of $f_{\eta'}^{u}$ $ f_{\eta'}^{s}$ or alternately
the values of $f_1$ and $f_8$, $N_c$ and the mixing angle. 
We will try to fit the
experimental number for $B \to K \eta'$ by assuming 
$f_1\sim (1.0 - 1.5) f_\pi$ and $f_8 \sim (1.0-1.5)f_\pi$.
We will take $N_c=2$, the effective number of colors. 
We find that there are several solutions corresponding to different
values for the set of parameters $f_1,f_8,\theta$ that  
can reproduce the experimental data on
$B \to K \eta'$ and the upper limit of $B \to K \eta $ if we use the
form factors in Ref\cite{Narduli} {\footnote { For the BSW model the
form factor $F_0(0)$ for $B \to \eta$ and $B \to \eta'$ are
approximately same and so we assume this  equality for the
form factors in Ref \cite{Narduli} where only the $ B \to \eta$ form factor is
calculated.}}
 For the CKM parameters we choose two sets
($\rho=0.15$ and $\eta=0.33$) and ($\rho=-0.15$ and $\eta=0.33$)
 \cite{ALI}. 
In Table.1  we
give the rates for the decays involving a $K^{(*)}$ and $\eta$, $\eta'$ and
$\pi$ in the final state 
for $ \theta =-17^0$  and $N_c =2$ for two sets of the CKM parameters
with various values of $f_1$ and $f_8$. Phenomenological studies
involving radiative decays of the $\eta$ and $\eta'$ indicate 
values for $f_1 \sim 1.1 f_{\pi}$ and $f_8\sim 1.3 f_{\pi}$ \cite{BALL}. We show the results of using the form factors in Ref \cite{Narduli} and
Ref \cite{BSW} in the third and fourth column of Table. 1. The entries in
Table. 1 correspond to the choice of CKM
parameters $(\rho=0.15,\eta=0.33)$ and for $\eta'$ in the final state we
also include a second
number in the parentheses corresponding to the second set of CKM parameters
 though the data seem to favor a positive value of $\rho$.
 The branching ratio for $B \to \eta K $ is
suppressed by about a factor of 10 or more relative to $B \to \eta' K $
while $B \to \eta K^* $ is enhanced relative to $B \to \eta' K^* $ by a
small amount. This is in qualitative agreement with Ref\cite{LIPKIN} but
we find a smaller enhancement 
$B \to \eta K^* $ relative to $B \to \eta' K^* $. 
From the branching ratios in Table. 1, it is not possible to rule out the
four quark operator explanation for the large branching ratio in 
$B^- \rightarrow \eta' K^-$. 
\begin{table}[thb]
\begin{center}
\begin{tabular}{|c|c|c|c|c|}
\hline
Process & Experimental BR \cite{CLEO1} & Branching Ratio \cite{Narduli} (BR) & BR \cite{BSW} & 
 $f_1, f_8 $ \\ 
\hline
                       &     &$5.33\times 10^{-5}$ & $1.05\times 10^{-5}$
                            & 1.0,1.0 \\ 
 
$B^-\rightarrow K^- \eta'$ &
$(7.1 ^{+2.5}_{-2.1} \pm 0.9)\times 10^{-5}$ &$6.1\times 10^{-5}$ & 
$1.32\times 10^{-5}$ & 1.1 , 1.3 \\

                        &    &$(6.44,7.83)\times 10^{-5}$ &
 $1.45\times 10^{-5}$   
                                    & 1.1 , 1.5 \\
\hline

                       &     &$4.97\times 10^{-5}$ & $9.82\times 10^{-6}$
                            & 1.0,1.0 \\ 
 
$B^0\rightarrow K^0 \eta'$ &
$(5.3 ^{+2.8}_{-2.2} \pm 1.2)\times 10^{-5}$ &$5.69\times 10^{-5}$ & 
$1.24\times 10^{-5}$ & 1.1 , 1.3 \\

                        &    &$(6.01,7.3)\times 10^{-5}$ &
 $1.36\times 10^{-5}$   
                                    & 1.1 , 1.5 \\
\hline
                       &     &$7.82\times 10^{-6}$ & $2.3\times 10^{-7}$
                            & 1.0,1.0 \\ 
 
$B^-\rightarrow K^- \eta$ &
$<8.0 \times 10^{-6}$ &$5.13\times 10^{-6}$ & 
$3.87\times 10^{-7}$ & 1.1 , 1.3 \\

                        &    &$(3.68,7.3)\times 10^{-6}$ & 
$7.5\times 10^{-7}$   
                                    & 1.1 , 1.5 \\ 
\hline
                       &     &$7.22\times 10^{-6}$ & $2.16\times 10^{-7}$
                            & 1.0,1.0 \\ 
 
$B^0\rightarrow K^0 \eta$ &
$<8.0 \times 10^{-6}$ &$4.72\times 10^{-6}$ & 
$3.68\times 10^{-7}$ & 1.1 , 1.3 \\

                        &    &$(3.39,6.7)\times 10^{-6}$ & 
$7.13\times 10^{-7}$   
                                    & 1.1 , 1.5 \\ 
\hline
&     &$1.18\times 10^{-5}$ & $4.79\times 10^{-7}$
                            & 1.0,1.0 \\ 
 
$B^-\rightarrow K^{-*} \eta'$ &
$<2.4 \times 10^{-4}$ &$1.15\times 10^{-5}$ & 
$5.67\times 10^{-7}$ & 1.1 , 1.3 \\

                        &    &$1.14\times 10^{-5}$ & $5.84\times 10^{-7}$   
                                    & 1.1 , 1.5 \\ 
\hline
&     &$1.36\times 10^{-5}$ & $9.43\times 10^{-7}$
                            & 1.0,1.0 \\ 
 
$B^-\rightarrow K^{-*} \eta $ &
$<2.4 \times 10^{-4}$ &$1.31\times 10^{-5}$ & 
$7.38\times 10^{-7}$ & 1.1 , 1.3 \\

                        &    &$1.27\times 10^{-5}$ & 
$6.20\times 10^{-7}$   
                                    & 1.1 , 1.5 \\ 
\hline                     
$B^{\pm}\rightarrow K \pi^{\pm}$ &
$(2.3 ^{+1.1 +0.2}_{-1.0-0.2}\pm 2.0)\times 10^{-5}$ 
&$2.25 \times 10^{-5}$ & 
$8.87\times 10^{-6}$ & -- \\
\hline
$B^{\pm}\rightarrow K^* \pi^{\pm}$ &
$----$ 
&$1.50 \times 10^{-5}$ & 
$5.92\times 10^{-6}$ & -- \\
\hline
 $B^0({\bar B}^0)\rightarrow K^{\mp} \pi^{\pm}$ &  $(1.5^{+0.5+0.1}_{-0.4-0.1}\pm
0.1)\times 10^{-5}
$ &
$2.21\times 10^{-5}$& $8.79\times 10^{-6}$   
& --\\
\hline
 $B^0({\bar B}^0)\rightarrow K^{*\mp} \pi^{\pm}$ &  $--
$ &
$1.21\times 10^{-5}$& $4.77\times 10^{-6}$   
& --\\
\hline
 $B^{\pm}\rightarrow K^{\pm} \pi^0 $& 
$ (1.6^{+0.6+0.3}_{-0.5-0.2} \pm 0.1) \times 10^{-4}$ &$1.02\times 10^{-5}$ 
 & $ 4.25 \times 10^{-6} $ 
& -- \\
\hline
$B^{\pm}\rightarrow K^{\pm*} \pi^0$ &----- &$5\times 10^{-6}$ 
 & $ 2.3 \times 10^{-6} $ & -- \\
\hline 
\end{tabular}
\end{center}
\label{Tb_integrated}
\end{table}

For the quasi-inclusive decay we will use the same parameters as used in
the exclusive decays with the second set of the CKM parameters as this set
gives the larger rate between the two choices. We plot
the decay distribution $d\Gamma/dM_{rec}$ in Fig. 1 showing the
contributions from the effective Hamiltonian of four quark operators. 
From Fig. 1 we find the 
contribution from
the four quark operators to the branching ratio
$B\to \eta' X$ is around $1.3 \times 10^{-4}$ for
$E_{\eta'}>2.2 $ which is the signal region. This is far too small
to account for the signal observed at high $X_s$ mass. This continues to
be true even if when all parameters are allowed to vary over a
reasonable range.  

\section{Conclusion}
Summarizing, we have calculated the the effects of the effective
Hamiltonian of four quark operators to the exclusive and quasi-inclusive
decays of the B meson to $\eta'$. Our analysis indicate that the
 exclusive data can be explained by the four quark operators without
significant contribution from
 the mechanism $b \to s g^*$, $ g^* \to \eta' g g$ or from the
intrinsic charm content of the $\eta^{\prime}$. The contribution to the
quasi-inclusive rate from the four quark operator is not enough to
account for the observed signal.
\section{Acknowledgments}
 This work was
supported by the United States Department of Energy under contracts
DE-FG 02-92ER40730 and DE-FG 03-94ER40833 and by the Australian Research 
Council. We thank T.E. Browder
for useful discussions.

\section{Figure Caption}
\begin{itemize}
\item{\bf Fig. 1} This figure shows the decay distribution as a function
of the recoil mass $M_{rec}$.
\end{itemize}
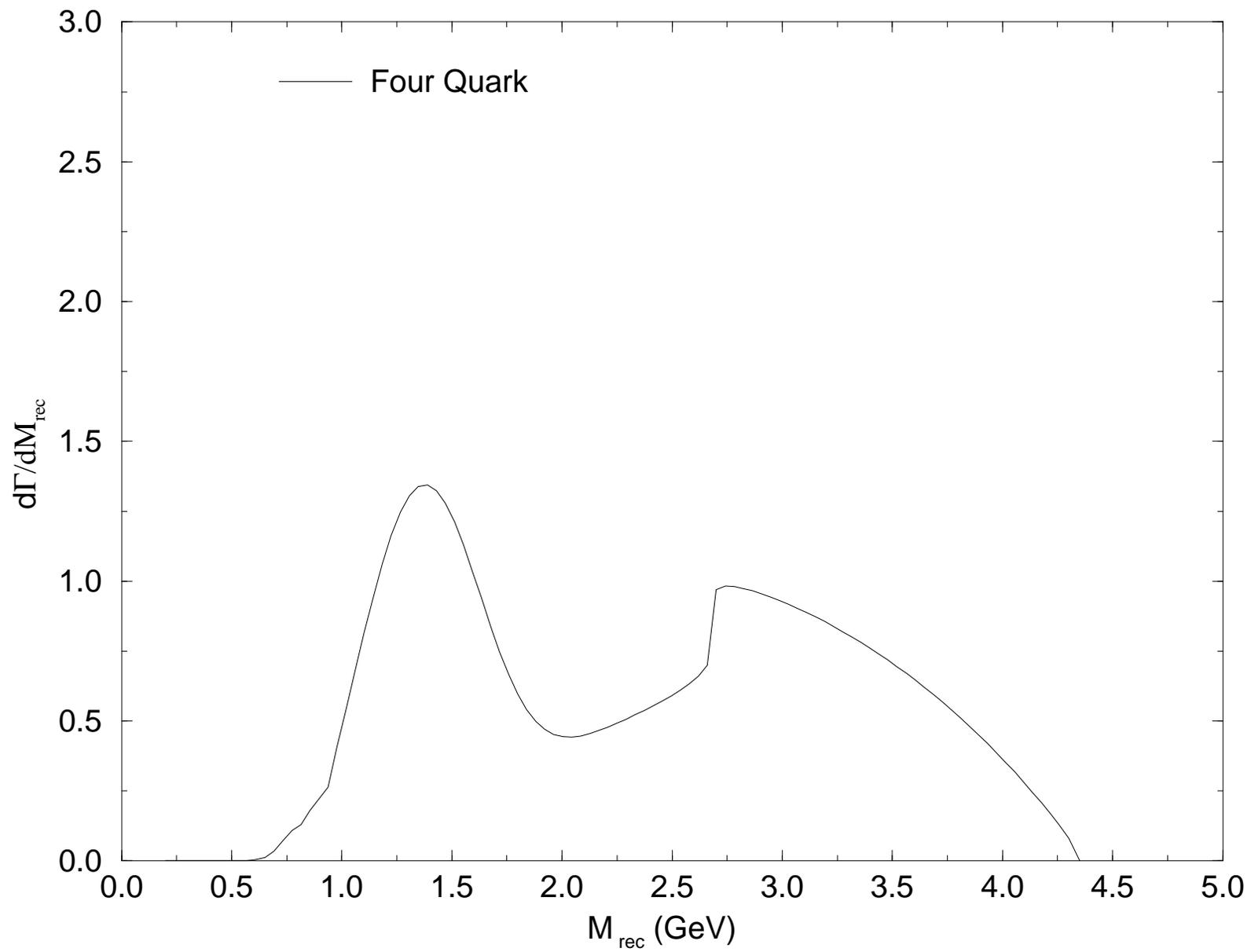
\end{document}